\documentclass[aps,prl,twocolumn,epsfig]{revtex4}

\usepackage{graphicx}
\usepackage{amsmath}
\usepackage{epsfig,times}
\usepackage{dcolumn}
\usepackage{bm}
\usepackage{color}
\usepackage{epstopdf}
\usepackage{amssymb}
\usepackage{amstext}
\usepackage{latexsym}
\usepackage{hyperref}

\usepackage{amsfonts}

\newcommand{\trace}{{\rm Tr}}
\newcommand{\beq}{\begin{equation}}
\newcommand{\eeq}{\end{equation}}
\newcommand{\bea}{\begin{eqnarray}}
\newcommand{\eea}{\end{eqnarray}}

	\newcommand{\tr}[1]{\textrm{Tr} \left[ {#1} \right]} 

	\newcommand{\e}[1]{e^{ {#1}}} 

	\newcommand{\s}[2]{\sigma^{#1}_{#2}} 

	\newcommand{\integral}[3]{{\int^{#2}_{#3} \mathrm{d}{#1} \;}} 

	\newcommand{\be}{\begin{equation}}
	\newcommand{\ee}{\end{equation}}

	\newcommand{\ket}[1]{{\left\vert {#1} \right\rangle}}	
	\newcommand{\bra}[1]{{\left\langle {#1} \right\vert}}

\begin{document}

\title{Emergent thermodynamics in a quenched quantum many-body system}

\author{R.~Dorner$^{1,2}$}
\email{rd309@ic.ac.uk}
\author{J.~Goold$^{2,3}$}
\author{C.~Cormick$^{4}$}
\author{M.~Paternostro$^{5,6}$}
\author{V.~Vedral$^{2,7,8}$}
\affiliation{$^1$Blackett Laboratory, Imperial College London, London SW7 2AZ, United Kingdom}
\affiliation{$^2$Clarendon Laboratory, University of Oxford, Oxford OX1 3PU, United Kingdom}
\affiliation{$^3$ Physics Department, University College Cork, Cork, Ireland}
\affiliation{$^4$ Theoretische Physik, Universit\"{a}t des Saarlandes, D-66123 Saarbr\"{u}cken, Germany}
\affiliation{$^5$Centre for Theoretical Atomic, Molecular and Optical Physics, School of Mathematics and Physics, Queen's University Belfast, BT7 1NN Belfast, United Kingdom}
\affiliation{$^6$Institut f\"ur Theoretische Physik, Albert-Einstein-Allee 11, Universit\"{a}t Ulm, D-89069 Ulm, Germany}
\affiliation{$^7$ Center for Quantum Technologies, National University of Singapore, 3 Science Drive 2, 117543 Singapore, Singapore}
\affiliation{$^8$ Department of Physics, National University of Singapore, 2 Science Drive 3, 117542 Singapore, Singapore}

\date{\today}

\begin{abstract}
We study the statistics of the work done, the fluctuation relations and the irreversible entropy production in a quantum many-body system subject to the sudden quench of a control parameter.
By treating the quench as a thermodynamic transformation we show that the emergence of irreversibility in the nonequilibrium dynamics of closed many-body quantum systems can be accurately characterized. 
We demonstrate our ideas by considering a transverse quantum Ising model that is taken out of equilibrium by the instantaneous switching of the transverse field.
\end{abstract}
\maketitle

\noindent
{\it Introduction--}
In the past decade or so, there has been a revival of interest in the study of nonequilibrium dynamics in closed quantum systems. Mainly, this is due to a series of spectacular experiments using ultracold atoms, where the high degree of isolation and long coherence times permit the study of dynamics over long timescales \cite{Greiner2:02}. These experiments have raised a number of important theoretical issues including the relationship between thermalisation and integrability \cite{Polkovnikov:11} and the universality of defect generation in the adiabatic crossing of a critical point \cite{dziamaga}. A common way to take a many-body system out of equilibrium is by an abrupt change of a local or global parameter of the Hamiltonian, this is commonly referred to as a `sudden quench'. 
Following a quench the dynamical response of the system can be probed by studying, for example, the dynamical correlation functions~\cite{cardy}, change in the diagonal entropy~\cite{polkinov} or the statistics of work done~\cite{Silva}. 

Over a similar period of time, there has also been a great deal of interest in the statistical mechanics community surrounding the discovery of the nonequilibrium fluctuation relations (see e.g. Ref.~\cite{jrev} for a review).
Essentially, the fluctuations relations encode the full non-linear response of a system to a time dependent change of a Hamiltonian parameter.
In particular, they make a definitive statement regarding the irreversible entropy production of a system following a thermodynamic transformation and, as such, allow us to understand the emergence of thermodynamic behaviour in systems where the microscopic laws are inherently reversible. 
Given the current experimental interest in the nonequilibrium dynamics of ultra-cold atomic systems 
and the recent developments in statistical mechanics, it is natural to study the quench dynamics of quantum many-body systems in this new thermodynamical formulation. 
In this work we use the transverse Ising model~\cite{Karevski} to provide an exact analysis of the Tasaki-Crooks and Jarzynski fluctuation relations in a quenched many-body system. Furthermore, we compute the average irreversible production and show that emergence of thermodynamics provides an elegant interpretation of the essential physics.

\noindent
{\it Nonequilibrium quantum thermodynamics--}
We begin by reviewing some key concepts of microscopic thermodynamics, allowing us to define the formalism that is used in the rest of our study.

One of the fundamental goals of quantum thermodynamics is to understand how thermodynamical laws emerge from the underlying quantum mechanics of individual particles~\cite{mahlerbook}.
In this spirit, we consider a dynamical system described by a Hamiltonian $H(\lambda(t))$ that depends on an external work parameter $\lambda(t)$, i.e. an externally controlled parameter whose value determines the equilibrium configuration of the system.
The system is prepared by allowing it to equilibrate with a heat reservoir at inverse temperature $\beta$ for a fixed value of the work parameter $\lambda(t\leq0)=\lambda_0$. The initial state of the system is thus the Gibbs state
$\rho_\textrm{G}(\lambda_0)$, where
\begin{equation}
\rho_\textrm{G}(\lambda):=\frac{\e{-\beta H(\lambda)}}{\mathcal{Z}(\lambda)},
\nonumber
\end{equation}
and the partition function $\mathcal{Z}(\lambda)=\tr{\e{-\beta H(\lambda)}}$. At $t=0$ the resevoir-system coupling is removed and a protocol is performed on the system taking the work parameter from its initial value $\lambda_0$ to a final value $\lambda_\tau$ at a later time $t=\tau$.
The initial and final Hamiltonians connected by the protocol $\lambda_0\to\lambda_\tau$ have the spectral decompositions $H(\lambda_0)=\sum_n \epsilon_n(\lambda_0) \ket{n}\bra{n}$ and $H(\lambda_\tau)=\sum_m \epsilon'_m(\lambda_\tau) \ket{m}\bra{m}$, respectively,
where $\ket{n}$ ($\ket{m})$ is the $n^\textrm{th}$ ($m^\textrm{th}$) eigenstate of the initial (final) Hamiltonian with eigenvalue $\epsilon_n$ ($\epsilon'_m$).

The definition of the work done on the system $W$ as a consequence of the protocol requires two projective measurements; The first projects onto the eigenbasis of the initial Hamiltonian
$H(\lambda_{0})$ at $t=0$, with the system in thermal equilibrium. 
The system then evolves under the unitary dynamics $U(\tau; 0)$ generated by the protocol $\lambda_0\to\lambda_\tau$ before the second measurement
projects onto the eigenbasis of the final Hamiltonian $H(\lambda_\tau)$. The probability of obtaining
$\epsilon_n$ for the first measurement outcome followed by $\epsilon'_m$ for the second measurement is then $p_n^0p_{m|n}^\tau=\e{-\beta \epsilon_n}|\bra{n}U(\tau,0)\ket{m}|^2/\mathcal{Z}(\lambda_0)$. Accordingly, the
work distribution is defined as~\cite{Tasaki}
\begin{equation}
P_\textrm{F}(W):=\sum_{n,m} p^0_n\;  p^\tau_{m \vert n} \delta\left[W-(\epsilon_m'-\epsilon_n)\right].
\label{eq:qworkdist}
\end{equation}
Eq.~\eqref{eq:qworkdist} therefore encodes the 
fluctuations in the work that arise from thermal statistics
($p_n^0$) and from quantum measurement statistics ($p^\tau_{m \vert n}$) over many identical realisations of the protocol.
For our purposes, it is convenient to define the characteristic function of the work distribution as the Fourier transform of Eq.~\eqref{eq:qworkdist}~\cite{lutz}
\begin{align}
\chi_F(u,\tau) &= \integral{W}{}{}\e{iuW}P_\textrm{F}(W),
\nonumber \\
&=\tr{U^\dag(\tau,0)\e{iuH(\lambda_\tau)}U(\tau,0)\e{-iuH(\lambda_0)}\rho_\textrm{G}(\lambda_0)}.
\label{eq:loschmidt}
\end{align}
The convenience of $\chi_\textrm{F}(u,\tau)$ is evident when considering the well-known Tasaki-Crooks fluctuation relation $P_\textrm{F}(W)/P_\textrm{B}(-W)=\e{\beta(W-\Delta F)}$~\cite{Crooks,Tasaki}. This states that the ratio between the {\it forward} work distribution $P_\textrm{F}(W)$, introduced above, and the {\it backward} work distribution $P_\textrm{B}(-W)$, obtained from the protocol $\lambda_\tau\to \lambda_0$ in which the system is initialized at $t=0$ in the Gibbs state $\rho_\textrm{G}(\lambda_{\tau})$ and evolves according to $U^\dag(\tau,0)$, is related to the difference in the equilibrium free-energy of the system $\Delta{F}$.
Following Ref.~\cite{crooks2page} the Tasaki-Crooks relation is written in terms of the characteristic function as 
\begin{equation}
\frac{\chi_F(u,\tau)}{\chi_{B}(v,\tau)}=\frac{\mathcal{Z}(\lambda_\tau)}{\mathcal{Z}(\lambda_0)},
\label{eq:recluse}
\end{equation}
where
we have introduced the backward characteristic function
$\chi_B(v) = \integral{W}{}{}\e{ivW}P_\textrm{B}(W)$.
Moreover, the Jarzynski equality~\cite{Jarzynski} is easily obtained from Eq.~\eqref{eq:loschmidt} by introducing the parameter $u=i\beta$, giving
\begin{equation}
\chi_\textrm{F}(i\beta,\tau)= \langle \e{-\beta W} \rangle=\frac{\mathcal{Z}(\lambda_\tau)}{\mathcal{Z}(\lambda_0)}=\e{-\beta \Delta F},
\label{eq:Jarzynski}
\end{equation}
where in obtaining the last equality we have used the relation $\Delta F= -(1/\beta)\ln\left(\mathcal{Z}(\lambda_\tau)/\mathcal{Z}(\lambda_0)\right)$.
Both the Tasaki-Crooks and Jarzynski fluctuation relations are statements regarding the symmetry of fluctuations in work during thermodynamic transformations of microscopic systems. Remarkably, these symmetries are solely determined by the equilibrium state quantity $\Delta F$ regardless of how far the system is driven from equilibrium. For a recent information-theoretic interpretation of the fluctuation relations see Ref.~\cite{infoth}.

\noindent
{\it Irreversible entropy production--}
For finite systems, the statistical nature of work Eq.~\eqref{eq:qworkdist} requires the second law of thermodynamics to be revised to the form
$\langle W \rangle\ge\Delta{F}$,
with equality being reached for a quasistatic process. For all non-ideal processes, the deficit between average work $\langle W \rangle$ and the variation in free energy can be accounted for  by the {\it ad hoc} introduction of the average irreversible work,
\begin{equation}
\langle W \rangle = \langle W_\textrm{irr} \rangle+\Delta F.
\nonumber
\end{equation}
For a closed quantum system, the heat transfer into the system $\langle Q \rangle=0$ and the sole contribution to the change in entropy is the average irreversible entropy production $\Delta S_\textrm{irr}$.
In Ref.~\cite{lutz2} it is shown that for an initial Gibbs state $\rho_\textrm{G}(\lambda_0)$ undergoing unitary evolution generated by a time dependent Hamiltonian $H(\lambda(t))$, the average irreversible entropy production is given by the relative entropy between the instantaneous state of the system $\rho(t)=U(t,0)\rho_\textrm{G}(\lambda_0)U^\dag(t,0)$ and a hypothetical Gibbs state at that time,
\begin{align}
\langle \Delta S_\textrm{irr} \rangle&=S\left(\rho(t)||\rho_\textrm{G}(\lambda_t)\right)
\nonumber \\
&= \tr{\rho(t)\log\rho(t)}-\tr{\rho(t)\log \rho_\textrm{G}(\lambda(t))}.
\label{eq:lutzentropy}
\end{align}
In the case of a sudden quench, in which the work parameter $\lambda(t)$ is suddenly switched between some initial and final value, Eq.~\eqref{eq:lutzentropy} becomes
\beq
\langle \Delta S_\textrm{irr} \rangle=S(\rho_\textrm{G}(\lambda_0)||\rho_\textrm{G}(\lambda_\tau))=\beta(\langle W\rangle-\Delta F)=\beta\langle W_\textrm{irr}\rangle.
\label{eq:irrworkent}
\eeq
This expression for the average irreversible entropy production induced by a sudden quench in a closed quantum system was first 
discovered by Donald in Ref.~\cite{Donald} within a different context.

\noindent
{\it Transverse Ising model--}
We now apply the framework of nonequilibrium statistical mechanics outlined above to the nonequilibrium transformation of a thermal quantum spin chain.
In particular, we analyse the sudden quench of the transverse field in the quantum Ising model. For a discussion of this model in the zero temperature limit see Refs.~\cite{Silva,Dziarmagaprl}.

We consider a one-dimensional ring of $N$ spin-$1/2$ particles that interact with their nearest-neighbours via ferromagnetic coupling along the $z$-axis and with an external field applied along the $x$-axis. The Hamiltonian is
\begin{equation}
H = - \sum_{j=1}^N \lambda \sigma^x_j + \sigma^z_j \sigma^z_{j+1},
\label{eq:Ising1}
\end{equation}
where $\lambda$ is a dimensionless parameter measuring the strength of the external field with respect to the spin-spin coupling, $\sigma^{\alpha}_j$ ($\alpha=x,y, z$) is the spin-$1/2$ Pauli operator  acting at the $j^\textrm{th}$ spin and periodic boundary conditions are imposed by requiring that ${\sigma}^\alpha_{N+1}={\sigma}^\alpha_1$. 
The Hamiltonian Eq.~\eqref{eq:Ising1} is diagonalised by decomposing the Hilbert space into orthogonal parity subspaces and following the procedure outlined in the supporting material. In this way, considering the positive parity subspace only, the initial Hamiltonian with $\lambda=\lambda_0$ is written~\footnote{Here we have abused notation slightly for the sake of brevity. The Hamiltonian in Eq.~\eqref{eq:initialdiagham} is the positive parity contribution to the full Hamiltonian Eq.~\eqref{eq:Ising1} only. In the main text this fact this is indicated by the sum over the set of positive parity psuedomomenta $K^+$. In the supporting material, where the full diagonalisation of Eq.~\eqref{eq:Ising1} is discussed, the positive and negative parity contributions to the Hamiltonian are explicitly denoted $H^\pm$.}
\begin{equation}
H(\lambda_0)=\sum_{k \in K^+}\epsilon_k(\lambda_0)  \left(\gamma^\dagger_k \gamma_k -\frac{1}{2}\right),
\label{eq:initialdiagham}
\end{equation}
where $\gamma_k, \gamma^\dag_k$ are fermionic creation and annihilation operators labelled by the members of the set $K^+=\left\{\pm {\pi}(2n-1)/N: n=1,...{N}/{2}\right\}$ of {\it positive parity subspace} pseudomomenta.
Proceeding as earlier, the system is prepared in the Gibbs state $\rho_\textrm{G}(\lambda_0)=\e{-\beta H(\lambda_0)}/\mathcal{Z}(\lambda_0)$ with inverse temperature $\beta$ and associated partition function
\begin{equation}
\mathcal{Z}(\lambda_0)=
2^N \prod_{\substack{k \in K^+\\k>0}} \textrm{cosh}^2\left(\frac{\beta \epsilon_k(\lambda_0)}{2} \right).
\nonumber
\end{equation}
The protocol constitutes the instantaneous switching of the transverse field to the final value $\lambda=\lambda_\tau$, giving the final Hamiltonian
\begin{equation}
H(\lambda_\tau)=\sum_{k \in K^+} \epsilon(\lambda_\tau) \left(\tilde{\gamma}^\dagger_k \tilde{\gamma}_k -\frac{1}{2}\right).
\label{eq:finaldiagham}
\end{equation}
Note that the differing values of the transverse field in Eqs.~\eqref{eq:initialdiagham} and \eqref{eq:finaldiagham} require diagonalising transformations that are quantitatively different. Consequently the post-quench fermionic operators $\{\tilde{\gamma}_k\}$ differ from their pre-quench counterparts $\{\gamma_k\}$, though the allowed values of the psuedomomenta are identical in both cases.
In the case of a sudden quench the characteristic function Eq.~\eqref{eq:loschmidt} takes the simplified form
\begin{equation}
\chi_\textrm{F}(u)=\tr{\e{iuH(\lambda_\tau)}  \e{-iuH(\lambda_0)} \rho_\textrm{G}^+(\lambda_0)}.
\label{eq:isingcharfn}
\end{equation}
Using Eqs.~\eqref{eq:initialdiagham} and \eqref{eq:finaldiagham}, the trace in Eq.~\eqref{eq:isingcharfn} is taken over the eigenstates of the initial Hamiltonian $\left\{\ket{n_k,n_{-k}}\right\}$ to give
\begin{align}
\chi_\textrm{F}(u)={}&\frac{1}{\mathcal{Z}(\lambda_0)}\prod_{\substack{k \in K^+\\k>0}} \sum_{n_{\pm k}=0,1}  \e{-(iu+\beta)\epsilon_k(\lambda_0)(n_k+n_{-k}-1)}
\nonumber \\
&\times \bra{n_k,n_{-k}}\e{iu\epsilon_k(\lambda_\tau) \left(\tilde\gamma^\dagger_k \tilde\gamma_k +\tilde\gamma^\dagger_{-k} \tilde\gamma_{-k}-1\right)}\ket{n_k,n_{-k}}.
\nonumber
\end{align}
The matrix elements can be evaluated explicitly to give an analytic form of the forward characteristic function. Hence,
%
%
%
\begin{align}
\chi_\textrm{F}(u)={}&\frac{1}{\mathcal{Z}(\lambda_0)}\prod_{\substack{k \in K^+\\k>0}} \e{\left(iu+\beta\right)\epsilon_k(\lambda_0)}\left(C_k^-(u,\lambda_\tau)+S_k^+(u,\lambda_\tau)\right)
\nonumber \\
&+\e{-\left(iu+\beta\right)\epsilon_k(\lambda_0)}\left(C_k^+(u,\lambda_\tau)+S_k^-(u,\lambda_\tau)\right)+2.
\label{eq:qfwdchareq}
\end{align}
Here we have introduced the quantities
$C_k^\pm(u,\lambda)=\cos^2 \left(\Delta_k/2 \right)\e{\pm iu\epsilon_k(\lambda)}$
and $S_k^\pm(u,\lambda)=\sin^2 \left(\Delta_k/2 \right)\e{\pm iu\epsilon_k(\lambda)}$, where $\Delta_k = \tilde\phi_k-\phi_k$ is the difference in the pre- and post-quench Bogolyubov angles (see supporting information). 

\noindent
{\it Verification of the fluctuation theorems--}
The verification of the Tasaki-Crooks relation Eq.~\eqref{eq:recluse} requires an expression for the backward characteristic function. This is easily obtained using a procedure similar to that described above for the forward characteristic function Eq.~\eqref{eq:qfwdchareq} under the mapping
$\lambda_0 \leftrightarrow\lambda_\tau$ $\Rightarrow$ $\epsilon_k(\lambda_0) \leftrightarrow\epsilon_k(\lambda_\tau)$, $\Delta_k \to-\Delta_k$.
The Tasaki-Crooks relation follows from $\chi_\textrm{B}(v)$ by introducing the complex parameter $v=-u+i\beta$. Noting that $C_k^\pm(-u+i\beta,\lambda)=C_k^{\mp}(u,\lambda)\e{\mp\beta\epsilon_k(\lambda)}$ and $S_k^\pm(-u+i\beta,\lambda)=S_k^{\mp}(u,\lambda)\e{\mp\beta\epsilon_k(\lambda)}$, it is straightforward to show that
\begin{align}
\chi_\textrm{B}(v)={}&\frac{1}{\mathcal{Z}(\lambda_\tau)}\prod_{\substack{k \in K^+\\k>0}}
\e{\left(iu+\beta\right)\epsilon_k(\lambda_0)}\left(C_k^-(u,\lambda_\tau)+S_k^+(u,\lambda_\tau)\right)
\nonumber \\
&+\e{-\left(iu+\beta\right)\epsilon_k(\lambda_0)}\left(C_k^+(u,\lambda_\tau)+S_k^-(u,\lambda_\tau)\right)+2.
\nonumber
\end{align}
The ratio of the forward and backward characteristic functions is thus
\begin{equation}
\frac{\chi_\textrm{F}(u)}{\chi_\textrm{B}(v)}=\frac{\mathcal{Z}(\lambda_\tau)}{\mathcal{Z}(\lambda_0)},
\nonumber
\end{equation}
which is equivalent to the Crooks relation Eq.~\eqref{eq:recluse}. Further, the Jarzynski equality Eq.~\eqref{eq:Jarzynski} follows from the forward characteristic function Eq.~\eqref{eq:qfwdchareq} by introducing the complex argument $u=i\beta$,
\begin{align}
\chi_\textrm{F}(i\beta)&=\frac{1}{\mathcal{Z}(\lambda_0)}\prod_{\substack{k\in K^+\\k>0}}\Big(2+2\textrm{cosh}\left(\beta\epsilon_k(\lambda_\tau)\right)\Big)
\nonumber \\
&=\frac{2^N}{\mathcal{Z}(\lambda_0)} \prod_{\substack{k\in K^+\\k>0}}\textrm{cosh}^2\left(\frac{\beta \epsilon_k(\lambda_\tau)}{2}\right)=\frac{\mathcal{Z}(\lambda_\tau)}{\mathcal{Z}(\lambda_0)}.
\nonumber
\end{align}
To our knowledge this is the first analytic demonstration of the fluctuation relations in a non-trivial quantum many-body system incorporating a critical point.

\noindent
{\it Emergent thermodynamics--}
The general form of the forward characteristic function following a sudden quench~ Eq.\eqref{eq:isingcharfn} admits a simple expression for the average work,
\beq
\label{eq:aws}
\langle W \rangle=\tr{H(\lambda_{\tau})\rho_\textrm{G}(\lambda_0)}-\tr{H(\lambda_{0})\rho_\textrm{G}(\lambda_0)}.
\eeq
Using the approach presented in the supporting material, the evaluation of Eq.~\eqref{eq:aws} leads to the following 
closed analytic form for the average work done on the spin system
%
\begin{align}
\label{analyticwork}
\langle W \rangle&=\sum_{\substack{k\in K^+\\k>0}}\left(\epsilon_{k}(\lambda_0)-\epsilon_{k}(\lambda_\tau)\cos(\Delta_k)\right)\tanh\left(\frac{\beta\epsilon_{k}(\lambda_0)}{2}\right)
\nonumber \\
&=2 \left(\lambda_0-\lambda_\tau \right) \sum_{\substack{k\in K^+\\k>0}} \cos\left(\phi_k\right)
\tanh\left(\frac{\beta\epsilon_k(\lambda_0)}{2}\right).
\nonumber
\end{align}
This in turn allows the calculation of the irreversible entropy production Eq.~\eqref{eq:irrworkent} for arbitrary temperature, spin number and quench amplitude,
\begin{align}
\langle \Delta S_\textrm{irr} \rangle=\beta\langle W \rangle +\sum_{\substack{k\in K^+\\k>0}} \ln\frac{\cosh^2\left(\beta \epsilon_k(\lambda_\tau\right)/2)}{\cosh^2\left(\beta \epsilon_k(\lambda_0)/2\right)}.
\nonumber
\end{align}

\begin{figure}
\centering
\includegraphics[width=\linewidth]{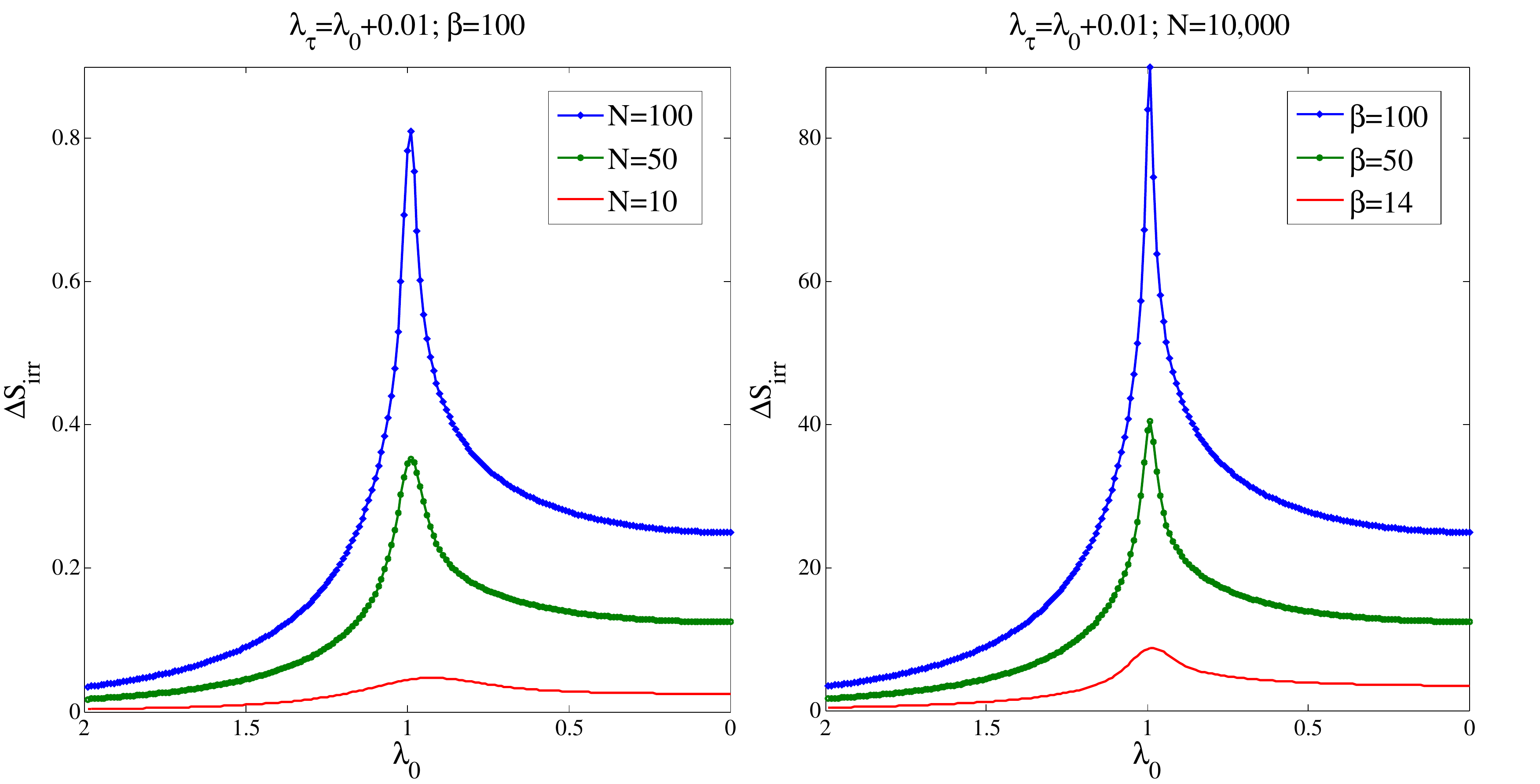}
\caption{Left: The irreversible entropy production for a series of quenches with amplitude $\vert\lambda_\tau-\lambda_0\vert=0.01$ at $\beta=100$ for several different ring sizes. The initial value of the transverse field $\lambda_{0}$ is shown on the $x$-axis. Right: The irreversible entropy production for the same series of quenches in a ring of $N=10,000$ spins at various temperatures.}
 \label{fig1}
\end{figure}

This quantity and its thermodynamic interpretation encapsulate the physics of the quench problem in a closed critical  system in a remarkably simple way. Fig.~\ref{fig1} shows the irreversible entropy production due to a series of sudden quenches with amplitude $\vert \lambda_\tau-\lambda_0\vert=0.01$. The left figure shows the quantity for spin chains of several sizes and low temperature $\beta=100$. The interpretation of the behaviour is straightforward; As the size of the system increases the energy gap between the ground and first excited state at the critical point begins to close. Work is performed to drive the system across the critical region and, due to the vanishing energy gap, it becomes increasingly difficult to do so without exciting the system, thereby dissipating work. This leads to the production of irreversible entropy and the emergence of intrinsic irreversibility in the critical region.

The figure on the right shows the irreversible entropy production in a chain of $N=10,000$ spins at various temperatures. 
As expected, the signature of quantum criticality decreases at higher temperatures with the emergence of thermal fluctuations.
The source of irreversibility is elucidated by manipulating the Tasaki-Crooks fluctuation relation to obtain the expression
$\langle \Delta S_\textrm{irr} \rangle=\integral{W}{}{}P_\textrm{F}(W) \log\left(P_\textrm{F}(W)/P_\textrm{B}(-W)\right)=\mathcal{K}\left(P_\textrm{F}(W)||P_\textrm{B}(W)\right)$,
where $\mathcal{K}\left(P_\textrm{F}(W)||P_\textrm{B}(W)\right)$ is the Kullback-Liebler relative entropy, measuring the distance between two probability distributions. Intuitively, this expression attributes the uncertainty in distinguishing the experimental data contained in the forward and backward work distributions to the degree of irreversible entropy production. According to this powerful interpretation, quantum criticality has the effect of setting the thermodynamic arrow of time as the irreversible entropy production grows with decreasing temperature.

As a final remark, we note that the forward characteristic function Eq.~\eqref{eq:loschmidt} has a similar form to the Loschmidt echo \cite{Loschmidt echo}. This has been shown in previous work to be a good indicator of phase transitions \cite{Quan} and could be experimentally measured using Raman interferometry
 \cite{Jens}.

\begin{acknowledgments}
{\it Acknowledgments --}
RD is funded by the EPSRC.
JG acknowledges funding from IRCSET through a
Marie Curie International Mobility fellowship; CC is supported by the Alexander von Humboldt Foundation; MP thanks the UK EPSRC for a Career Acceleration Fellowship and a grant under the ``New Directions for EPSRC Research Leaders" initiative (EP/G004579/1). VV is a fellow of Wolfson College Oxford and is supported by the John Templeton Foundation, the National Research Foundation and the Ministry of Education in Singapore, and the Leverhulme Trust (UK). JG is grateful to the Half Moon pub.  
\end{acknowledgments}

\renewcommand{\theequation}{S-\arabic{equation}}
\setcounter{equation}{0}  
\section*{Appendix A: Diagonalization of the transverse Ising model}  
\label{dia}

The quantum Ising model in a transverse field describes a lattice of spin-1/2 particles that interact
with their nearest-neighbours via ferromagnetic coupling along the $z$-axis and with an external
field applied along the $x$-axis. For a spatially homogeneous one-dimensional lattice of $N$ spins in a uniform field, the Hamiltonian is
\begin{equation}
H = - \sum_{j=1}^N \lambda \sigma^x_j + \sigma^z_j \sigma^z_{j+1},
\label{Aeq:Ising1}
\end{equation}
where $\lambda$ is a dimensionless parameter that measures the strength of the external field and the Pauli spin-$1/2$ operators are defined with periodic boundary conditions $\sigma^\alpha_{N+1}=\sigma^\alpha_1$ ($\alpha=x,y,z$). Under the canonical transformation $\sigma^x_j\to\sigma^z_j, \sigma^z_j \to -\sigma^x_j$ $\forall j$ the Hamiltonian Eq.~\eqref{Aeq:Ising1} becomes
\begin{equation}
H = - \sum_{j=1}^N \lambda \sigma^z_j + \sigma^x_j \sigma^x_{j+1},
\label{Aeq:Ising2}
\end{equation}
The spin operators are mapped to a spinless fermionic operators by a Jordan-Wigner transformation, thus
\begin{equation}
\begin{aligned}
c_j={}&\frac{1}{2}\prod_{l=1}^{j-1} \s{z}{l}(\s{x}{j}+i\s{y}{j}),
\nonumber \\
c_j^\dagger={}&\frac{1}{2}\prod_{l=1}^{j-1} \s{z}{l}(\s{x}{j}-i\s{y}{j}).
\end{aligned}
\nonumber
\end{equation}
Here, the operators $c_j$ ($c_j^\dagger$)  annihilate (create) a Jordan-Wigner fermion at the $j^\textrm{th}$ lattice site and obey the usual fermionic  anti-commutation relations. This in turn allows the definition of the parity operator
\begin{equation}
\Pi := \prod_{j=1}^N \left(1-2c_j^\dagger c_j \right),
\nonumber
\end{equation}
which measures whether the number of fermions in the chain is even ($\Pi=1$) or odd ($\Pi=-1$).
Following the Jordan-Wigner transformation the Hamiltonian Eq.~\eqref{Aeq:Ising2} factorises into two orthogonal parity subspaces,
\beq
H = P^+ H^+ P^+ + P^- H^- P^-.
\nonumber
\eeq
Here, $P^\pm = (1/2)\left(1\pm\Pi \right)$ are the projectors onto the even (+) and odd (-) parity subspaces and
\begin{equation}
H^{\pm}=-\lambda-\sum_{j}^{N} \left(2\lambda c_j^\dagger c_j-(c_j^\dagger c_{j+1}+c_{j+1} c_{j}+\textrm{h.c.})\right),
\label{eq:ising3}
\end{equation}
are the even and odd parity subspace contributions to the Hamiltonian. The even and odd parity subspace Hamiltonians are identical with the exception that in $H^+$ we impose the boundary condition $c_{N+1}=-c_1$ and in $H^{-}$ we impose $c_{N+1}=c_1$. Note that the parity of the chain is conserved ($\left[H^\pm,\Pi \right]=0$); the Hamiltonian in Eq.~\eqref{eq:ising3} does not mix the parity subspaces. Initialising the system in a state with zero projection onto, say, the odd subspace then restricts the dynamics to the even subspace only. For an initial Gibbs state, the system is a mixture of positive and negative parity states and both subspaces must be accounted for.
Despite this we restrict our attention to the even parity subspace only. This treatment becomes exact in the thermodynamic limit where boundary effects become negligible. This is the correct limit in which to discuss phase transitions, however a discussion of the fluctuation theorems is more suited to a finite chain. The analysis of a chain of arbitrary length in which both parity subspaces are accounted follows from a straightforward extension of what is presented here at the expense of more cumbersome expressions and provides little extra insight. Note that, in the main text, the fact that we consider the even subspace only is denoted by the summation over the set of positive parity pseudomomenta $k \in K^+$ in all relevant expressions. In the main text the `$+$' superscript that is used here to explicitly distinguish between the positive and negative parity subspaces is dropped for the sake of brevity.

The diagonalisation of the Hamiltonian is completed by application of a Fourier transformation followed by a Bogolyubov transformation, which factorizes over the spaces with different pseudomomentum $k$. For the positive parity contribution to the Hamiltonian the Fourier transformation is defined as
\beq
c_j = \frac{e^{-i\pi/4}}{\sqrt{N}} \sum_{k \in K^+} c_k e^{ikj},
\nonumber
\eeq
where the values of pseudomomentum are
\beq
K^+=\left\{k = \pm \frac{\pi}{N}(2n-1), \quad n=1, \ldots, \frac{N}{2}\right\}.
\nonumber
\eeq
After this transformation the Hamiltonian Eq.~\eqref{eq:ising3} takes the form
\beq
H^+ = \sum_{k \in K^+} 2\left(\lambda-\cos(k)\right) c^\dagger_kc_k + \sin(k) (c^\dagger_kc^\dagger_{-k} +c_{-k}c_k) -\lambda.
\nonumber
\eeq
Note that all the terms preserve pseudomomentum so that the remaining step of the diagonalization can be performed within each subspace with assigned value of $\pm k$. The last step is the Bogolyubov transformation
\begin{equation}
c_{\pm k}= \gamma_{\pm k} \cos\left(\frac{\phi_k}{2}\right) \mp \gamma^\dagger_{\mp k} \sin\left(\frac{\phi_k}{2}\right), 
\label{Aeq:bogrelzz}
\end{equation}
where
\begin{equation}
\begin{aligned}
\cos(\phi_k) &= \frac{\lambda-\cos(k)}{\sqrt{\sin^2(k) + [\lambda-\cos(k)]^2}},\\
\sin(\phi_k) &= \frac{\sin(k)}{\sqrt{\sin^2(k) + [\lambda-\cos(k)]^2}}.
\end{aligned}
\label{Aeq:bogangles}
\end{equation}
With this, the Hamiltonian can be written in the form
\beq
H^+ = \sum_{k \in K^+} \epsilon_k \left(\gamma^\dagger_k\gamma_k-\frac{1}{2}\right),
\nonumber
\eeq
with the dispersion relation
\beq
\epsilon_k = 2\sqrt{\sin^2(k) + [\lambda-\cos(k)]^2}.
\nonumber
\eeq
Note that $\epsilon_k = \epsilon_{-k}>0$ and that the total spectrum is symmetric with respect to the zero of energy.

\section*{Appendix B: Connecting the initial and final Hamiltonians}
To evaluate the characteristic function explicitly, the eigenstates of the initial Hamiltonian $H^+(\lambda_0)$ must be written in terms of the eigenstates of the final Hamiltonian $H^+(\lambda_\tau)$. Inverting Eq.~\eqref{Aeq:bogrelzz} and its hermitian conjugate it is possible to relate the sets of pre- and post-quench Bogolyubov operators. Hence,
\begin{equation}
\begin{aligned}
\tilde\gamma_k &= \gamma_k \cos\left(\frac{\Delta_k}{2}\right) + \gamma^\dagger_{-k} \sin\left(\frac{\Delta_k}{2}\right),
\nonumber \\
\tilde\gamma_{-k} &= \gamma_{-k} \cos\left(\frac{\Delta_k}{2}\right) - \gamma^\dagger_k \sin\left(\frac{\Delta_k}{2}\right).
\end{aligned}
\end{equation}
Here $\Delta_k = \tilde\phi_k - \phi_k$ and the expressions for the pre- and post-quench Bogolyubov angles, $\phi_k$ and $\tilde\phi_k$, have the form given in Eq.\eqref{Aeq:bogangles} with $\lambda=\lambda_0$ and $\lambda_\tau$ respectively.
Using this, the vacuum states in the two representations are related by
\begin{equation}
\ket{0_k,0_{-k}} = \left( \cos\left(\frac{\Delta_k}{2}\right) + \sin\left(\frac{\Delta_k}{2}\right) \tilde{\gamma}^\dagger_k \tilde{\gamma}^\dagger_{-k}\right) \ket{\tilde0_k,\tilde0_{-k}}.
\label{eq:gsrel}
\end{equation}
The expressions for higher energy eigenstates $\ket{n_k,n_{-k}}$ are then obtained by  applying the appropriate creation operators to Eq.~\eqref{eq:gsrel}. 

\section*{Appendix C: Calculation of the average work}
The calculation for the average work done on a quenched transverse Ising model takes advantage of the factorization of $H^+(\lambda_0)$ and $H^+(\lambda_0)$ into blocks of paired pseudomomenta with labels $\pm k$. We start by writing the density matrix of the system as $\rho=\bigotimes_{k>0}\rho_{\pm k}$, so that
\beq
\label{Aeq:aws1}
\langle W \rangle=\tr{\sum_{k>0}\left( H^+_{\pm k}(\lambda_\tau)-H^+_{\pm k}(\lambda_0)\right)\bigotimes_{k'>0} \rho_{\pm k'}},
\eeq  
where $ H^+_{\pm k}(\lambda_\tau) = \epsilon_k(\lambda_\tau)({\tilde\gamma_k}^\dagger\tilde\gamma_k + {\tilde\gamma_{-k}}^\dagger\tilde\gamma_{-k} -1)$ and
$ H^+_{\pm k}(\lambda_0) =  \epsilon_k(\lambda_0)({\gamma_k}^\dagger\gamma_k + {\gamma_{-k}}^\dagger\gamma_{-k} -1)$. With a little effort, Eq.~\eqref{Aeq:aws1} can be rewritten as
\beq
\label{Aeq:aws2}
\langle W \rangle=\sum_{k>0}\trace{\left[\left( H^+_{\pm k}(\lambda_\tau)-H^+_{\pm k}(\lambda_0)\right)\prod_{k'>0}\sigma_{\pm k'}\right]},
\eeq  
where,
\begin{equation}
\sigma_{\pm k}=\sum_{n_{\pm k}=0,1}|n_{k},n_{-k}\rangle\langle n_{k},n_{-k}|\frac{e^{-\beta\epsilon_{k}(\lambda_0)(n_{k}+n_{-k}-1)}}{4\cosh^{2}(\beta \epsilon_{k}(\lambda_0)/2)}.
\nonumber
\end{equation}
Noting that $\tr{\prod_{k'>0}\sigma_{\pm k'}}=1$, Eq.\eqref{Aeq:aws2} reduces to the form $\langle W\rangle=\sum_{k>0}\langle W_k\rangle$, with 
\beq
\langle W_{k}\rangle=\trace{\left[( H^+_{\pm k}(\lambda_\tau)-H^+_{\pm k}(\lambda_0)) \sigma_{\pm k}\right]}.
\nonumber
\eeq
In order to calculate the trace we note that we need only  keep the terms of $H^+_{\pm k}(\lambda_\tau)$ that are diagonal in the 
basis of $H^+_{\pm k}(\lambda_0)$;
\beq
\left[ H^+_{\pm k}(\lambda_\tau) \right]_\textrm{diag}=\epsilon_k(\lambda_\tau) \cos(\Delta_k)({\gamma_k}^\dagger\gamma_k + {\gamma_{-k}}^\dagger\gamma_{-k} -1).
\nonumber
\eeq
With this, $\langle W_{k} \rangle$ takes form
\beq
\langle W_k \rangle=\left(\cos(\Delta_k)\epsilon_{k}(\lambda_\tau)-\epsilon_{k}(\lambda_0)\right)\tr{(n_{k}+n_{-k}-1) \sigma_{\pm k}},
\nonumber
\eeq
which leads straightforwardly to the expression for the average work in the main text.

\end{document}